\documentclass[preprint,showpacs,preprintnumbers,amsmath,amssymb]{revtex4}

\usepackage{graphicx}
\usepackage{dcolumn}
\usepackage{bm}

\begin{document}

\preprint{APS/123-QED}

\title{Spin Singlet Formation from S = 1/2 Tetrahedral Clusters}

\author{T. Waki}
\affiliation{Department of Materials Science and Engineering, 
Kyoto University, Kyoto 606-8501, Japan}
\author{Y. Kajinami}
\affiliation{Department of Materials Science and Engineering, 
Kyoto University, Kyoto 606-8501, Japan}
\author{Y. Tabata}
\affiliation{Department of Materials Science and Engineering, 
Kyoto University, Kyoto 606-8501, Japan}
\author{M. Yoshida} 
\affiliation{Institute for Solid State Physics, University of Tokyo, Kashiwa 277-8581, Japan}
\author{M. Takigawa} 
\affiliation{Institute for Solid State Physics, University of Tokyo, Kashiwa 277-8581, Japan}
\author{I. Watanabe}
\affiliation{Advanced Meson Science Laboratory, RIKEN Nishina Center, Wako 351-0198, Japan}
\author{H. Nakamura}
\affiliation{Department of Materials Science and Engineering, 
Kyoto University, Kyoto 606-8501, Japan}
\date{\today}

\begin{abstract}
Muon spin relaxation ($\mu$SR) and nuclear magnetic resonance (NMR) experiments revealed that the spin singlet state with an excitation gap of $\sim$200 K is realized from $S = 1/2$ Nb$_4$ tetrahedral clusters in a cluster Mott insulator GaNb$_4$S$_8$.
The intercluster cooperative phenomenon to the singlet state at $T_\mathrm{S} = 32$ K is triggered by intracluster Jahn-Teller type structural instability developed from $\sim$$3T_\mathrm{S}$.
Referring to the lattice symmetry, the formation of Nb$_8$ octamer (Nb$_4$--Nb$_4$ bond) is suggested.
\end{abstract}


\maketitle

In any magnets, if spin is well defined at the atomic site, the classical spin degree of freedom
must be compensated on approaching 0 K, which is usually achieved by magnetic order.
Another route via electron-lattice coupling is quenching of spin itself by formation of a spin singlet state.
It is of interest to extend this concept to cluster compounds with mixed-valent magnetic clusters (more than one atom shares unpaired electrons within the cluster unit) embedded in a matrix crystal.
When the cluster unit has half-integer spin, particularly, since intracluster interactions never compensate all the spin, we expect long-range magnetic order due to minor intercluster interactions.
In this article, we show another possibility, formation of a new bound state between clusters, just like dimerization in the one-dimensional (1D) $S = 1/2$ spin chain with the electron-lattice coupling \cite{Peierls}.

\begin{figure}[b]
\includegraphics[width=0.9\linewidth]{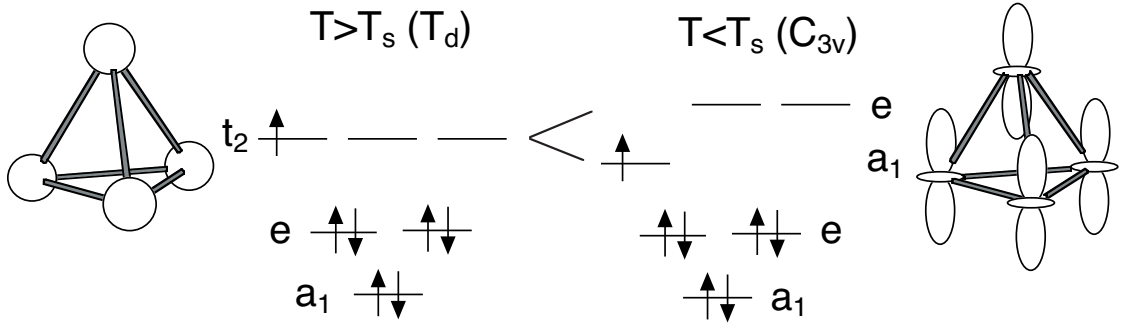}
\caption{\label{fig1}
Proposed molecular orbital schemes of the Nb$_4$ unit above and below $T_{\rm S}$ \cite{Pocha}. 
To interpret isotropic and anisotropic hyperfine fields at the Nb site, schematic spin density distributions are shown for HOMO.
}
\end{figure}

As a category of mixed-valent magnetic cluster compounds, ternary calcogenides AB$_4$X$_8$ (A = Ga, Al, Ge; B = V, Mo, Nb, Ta; X = S, Se) with the cubic GaMo$_4$S$_8$ structure (space group $F\bar{4}3m$) \cite{Brasen,BenYaich,Johrendt} are of particular interest.
In the compounds, cubic (B$_{4}$X$_{4}$)$^{n+}$ and tetrahedral (AX$_{4}$)$^{n-}$ ions are weakly coupled in a NaCl manner, resulting in hopping conduction between the clusters.
GaNb$_{4}$S$_{8}$ with (Nb$_{4}$S$_{4}$)$^{5+}$ and (GaS$_{4}$)$^{5-}$ ions has attracted attention as a sort of Mott insulator, where 4d electrons are localized at Nb$_4$ clusters due to electron correlations, and also as one of pressure-induced superconductors \cite{Pocha}.
 In molecular orbital schemes, the tetrahedral Nb$_4$ unit in (Nb$_{4}$S$_{4}$)$^{5+}$ shares seven 4d electrons, resulting in one unpaired electron at the highest occupied molecular orbital (HOMO) as shown in Fig.\ \ref{fig1} \cite{Pocha}.
 In other words, the Nb$_4$ tetramer has a local moment with $S = 1/2$.
Accordingly, the magnetic susceptibility $\chi$ obeys the Curie-Weiss law with the effective moment $\mu_\mathrm{eff}=1.73$ $\mu_\mathrm{B}$/f.u.\ \cite{Pocha}.
The negative Weiss constant $\theta=-298$ K
suggests antiferromagnetic (AF) coupling among clusters.
$\chi$ shows an abrupt drop at $T_\mathrm{S}=32$ K, which is accompanied by a structural transition to a tetragonal state ($P\bar{4}2_1m$) \cite{Jakob} possibly due to Jahn-Teller (JT) type instability; the triply degenerate HOMO splits to singlet and doublet states (see Fig.\ \ref{fig1}).
Concerning the magnetic ground state below $T_\mathrm{S}$, Pocha et al.\ \cite{Pocha} commented no additional reflections in powder neutron diffraction profiles, ruled out long-range AF order, and then ascribed the absence of long-range order to topological spin frustration.
On the other hand, in a more recent paper, Jacob et al.\ \cite{Jakob} proposed non-collinear AF order on the basis of their crystallographic analysis and band structure calculations.
Being inconsistent with $S = 1/2$ inherent to the cluster, 
our muon spin relaxation ($\mu$SR) and nuclear magnetic resonance (NMR) experiments clearly indicate presence of a spin excitation gap below $T_\mathrm{S}$, namely compensation of spin to the $S = 0$ state.
Referring to the lattice symmetry \cite{Pocha}, we discuss possible Nb$_8$ octamer formation from two Nb$_4$ tetramers.

All polycrystalline samples were prepared by solid state reaction from pure elements sealed in evacuated quartz tubes. 
Final heat treatment was performed at 900$^{\circ}$C for 1 day after several heat treatments and intermediate regrinds.
$\chi$, measured to check sample quality, was the same as literature data \cite{BenYaich,Jakob,Pocha} 
(our result is shown in Fig.\ \ref{fig3}.) 
Zero-field (ZF) and longitudinal-field (LF) $\mu$SR measurements were made at the RIKEN-RAL Muon Facility at the Rutherford-Appleton Laboratory in the UK using a pulsed positive surface muon beam at 4.5--100 K under LF fields of 0--0.4 T. 
NMR experiments for $^{69,71}$Ga (nuclear spin $I=3/2$ for both) and $^{93}$Nb ($I=9/2$) were performed with a standard phase coherent spectrometer.
Most NMR spectra were obtained by Fourier transform of the spin-echo or free induction decay signal in a fixed magnetic field of 7.0071 T.
The nuclear spin-lattice relaxation rate $1/T_1$ for $^{69,71}$Ga was obtained by inversion and saturation recovery methods.
$^{93}$Nb-$T_1$ is too short to be estimated reliably.

\begin{figure}[tb]
\includegraphics[width=0.9\linewidth]{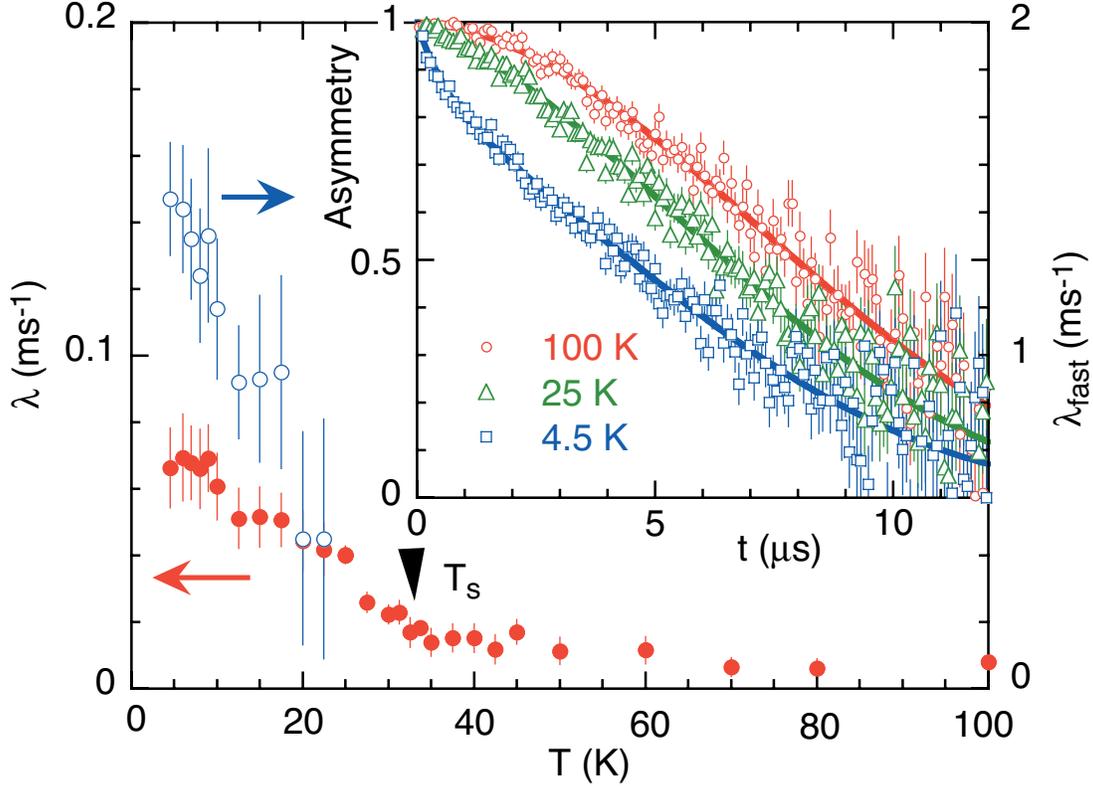}
\caption{\label{fig2}
(color online).
The $T$ dependence of the ZF muon spin relaxation rate $\lambda$. 
The inset shows typical examples of ZF-$\mu$SR spectra (at 4.5, 25 and 100 K). 
Solid curves indicate the fit by the damped KT function. 
Below $T_\mathrm{S}$, the asymmetry is fit together with another fast damping component.
}
\end{figure}

\begin{figure}[tbh]
 \begin{center}
 \includegraphics[width=0.9\linewidth]{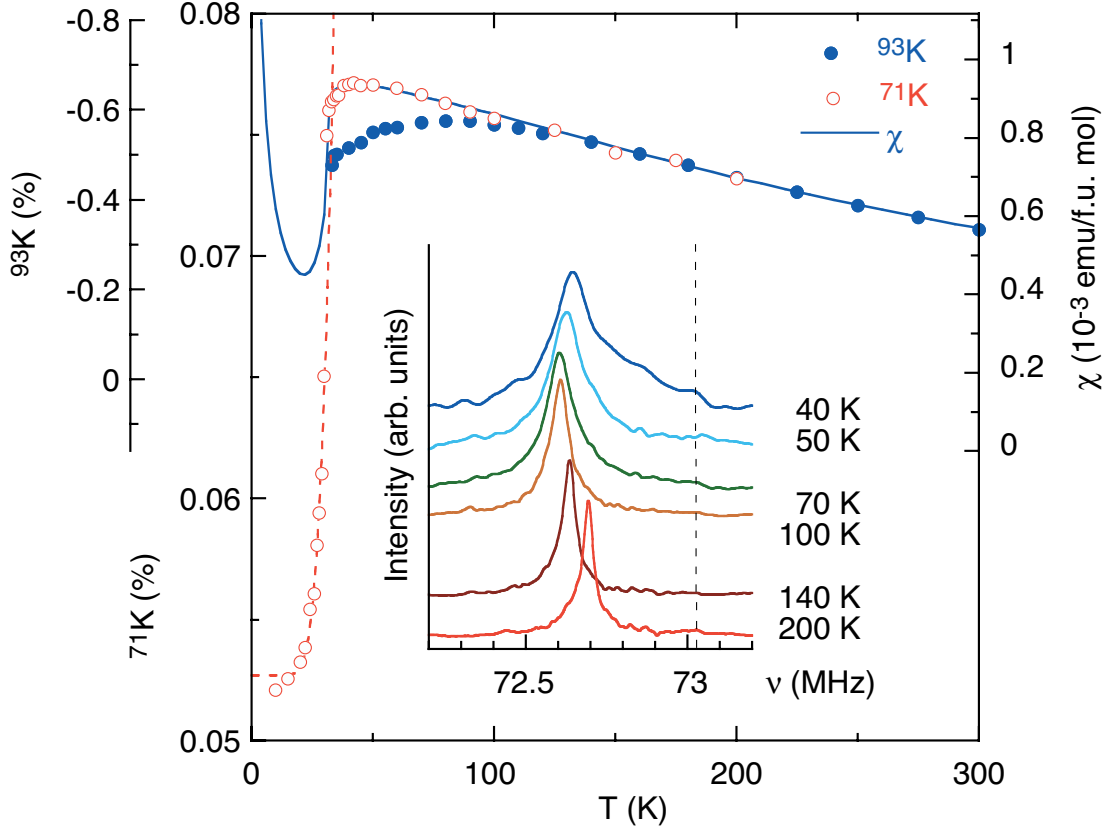}
 \end{center}
 \caption{\label{fig3} (color online).
$T$ dependences of $^{93}$Nb and $^{71}$Ga hyperfine shifts, $^{93}K$ (solid circles) and $^{71}K$ (open circles). 
$^{93}K$ is shown only above $T_\mathrm{S}$. 
$^{71}K$ below $T_\mathrm{S}$ corresponds to $^{71}K_\perp$ \cite{note1}.
The solid curve indicates the bulk susceptibility $\chi$.
The broken curve represents $\exp(-E_\mathrm{g}/k_\mathrm{B} T)$ with $E_\mathrm{g}/k_\mathrm{B} \simeq 220$ K.
The inset shows $^{93}$Nb-NMR lines (only the center line) above $T_\mathrm{S}$, measured under an external field of 7.0071 T. The broken straight line indicates the zero shift position. }
 \end{figure}

To determine the magnetic ground state, we first show results of $\mu$SR.
Since NMR is affected by both magnetic and electric hyperfine interactions, the analysis of powder-pattern spectra is often non-straightforward.
In contrast, $\mu$SR probes only the magnetic field, and has an advantage in monitoring the appearance of internal fields.
Typical examples of ZF-$\mu$SR spectra are shown in the inset of Fig.\ \ref{fig2}. 
Even at the lowest temperature ($T$), no evidence of magnetic order such as muon spin precession was observed. 
Above $T_\mathrm{S}$, the relaxation is dominated by Gaussian-like depolarization.
The commonly used damped Kubo-Toyabe (KT) function
$P_\mu(t) = \exp(-\lambda t) G_z^{\rm KT}(\Delta, t)$
with
$G_z^{\rm KT} (\Delta, t) = \frac{1}{3} + \frac{2}{3} (1 - \Delta^2 t^2) \exp (-\frac{1}{2} \Delta^2 t^2)$ 
was fit to the data, where $\Delta/\gamma_{\mu}$ is the width of the static field distribution ($\gamma_{\mu}$: muon gyromagnetic ratio), and $\lambda$ is the damping rate associated with an additional relaxation process mostly due to electron spin fluctuations. 
Above $T_\mathrm{S}$, by treating $\lambda$ and $\Delta$ as free fitting parameters, we obtained a nearly $T$-independent value of $\Delta \simeq 0.1$ $\mu \mathrm{s}^{-1}$, which corresponds to a tiny field of $\Delta/\gamma_{\mu} \simeq 1$ G, and is ascribed to dipolar fields coming from randomly oriented nuclear spins. 
With decreasing $T$ passing through $T_\mathrm{S}$, due to an appearance of fast relaxation, the above function tunes to be inappropriate.
Taking account of the symmetry lowering in the crystal, i.e., possible separation of the muon stopping site, we applied
$P_\mu(t) = (C-1) \exp(-\lambda t) G_z^{\rm KT}(\Delta, t) + C \exp(-\lambda _\mathrm{fast} t)$
to fit low-$T$ data, where $C$ is a constant.
In addition, associated with appearance of the fast relaxation, the damping part (dynamic fields) tends to mask the Gaussian part (small static fields), resulting in unreliable estimation of $\Delta$ below $\sim$30 K.
Then we assumed $T$ independence of $\Delta$ to analyze low-$T$ data.
Thus estimated $\lambda$ and $\lambda _\mathrm{fast}$ are plotted in Fig.\ \ref{fig2} ($C$ gradually increases with lowering $T$ and reaches $\sim$0.2 at the lowest $T$).
 $\lambda$ is enhanced strongly below $T_\mathrm{S}$ and approaches a constant value at the lowest $T$. 
This behavior is commonly seen for the system which condenses into a spin singlet state, and now understood as sporadic dynamics due to spin excitations in a singlet sea \cite{Uemura,Bono}.
In LF-$\mu$SR experiments below $T_\mathrm{S}$, strong muon-spin depolarization was observed in the whole range of LF up to 0.4 T. 
This observation indicates the presence of dynamic components of electronic spins and supports the absence of large static field due to static magnetic order.
Hence we conclude that the ground state is neither long-range nor short-range magnetic order.

Next let us see NMR data.
Examples of $^{93}$Nb-NMR lines measured above $T_\mathrm{S}$ are shown in the inset of Fig.\ \ref{fig3}, 
where only the center line (the $m = -1/2 \leftrightarrow 1/2$ transition) is shown.
Nearby the sharp center line, unresolved satellites due to quadrupole interaction, spreading out in several hundred kHz, were observed; the Nb atom occupies an axially symmetric site ($16e$) in the high-$T$ cubic state.
With decreasing $T$, the resonance frequency once decreases until $\sim$80 K and increases again.
The $T$ dependence of the magnetic hyperfine shift $^{93}K$, estimated from the peak position, is shown in Fig.\ \ref{fig3}. 
A high-$T$ part of $^{93}K$ scales well with $\chi$ \cite{note3}.
$|^{93}K|$ exhibits a characteristic broad hump at $\sim$80 K, deviating from a Curie-Weiss-like $T$ dependence at high $T$ \cite{note2}.
Furthermore it should be noted that the lineshape turns to be strongly asymmetric in spite of the reduction of $|^{93}K|$.
The origin of the characteristic hump and anisotropy of $^{93}K$ will be discussed later.
Below $T_\mathrm{S}$, the $^{93}$Nb spectrum turns to be broad and complex.
This is mainly due to separation of the Nb site to three crystallographically inequivalent sites ($8f$ and two $4e$) and appearance of large anisotropy in $^{93}K$.
Here we do not discuss further the low-$T$ $^{93}$Nb spectrum, because the analysis is not straightforward but complicated. 

\begin{figure}[tbh]
 \begin{center}
 \includegraphics[width=0.85\linewidth]{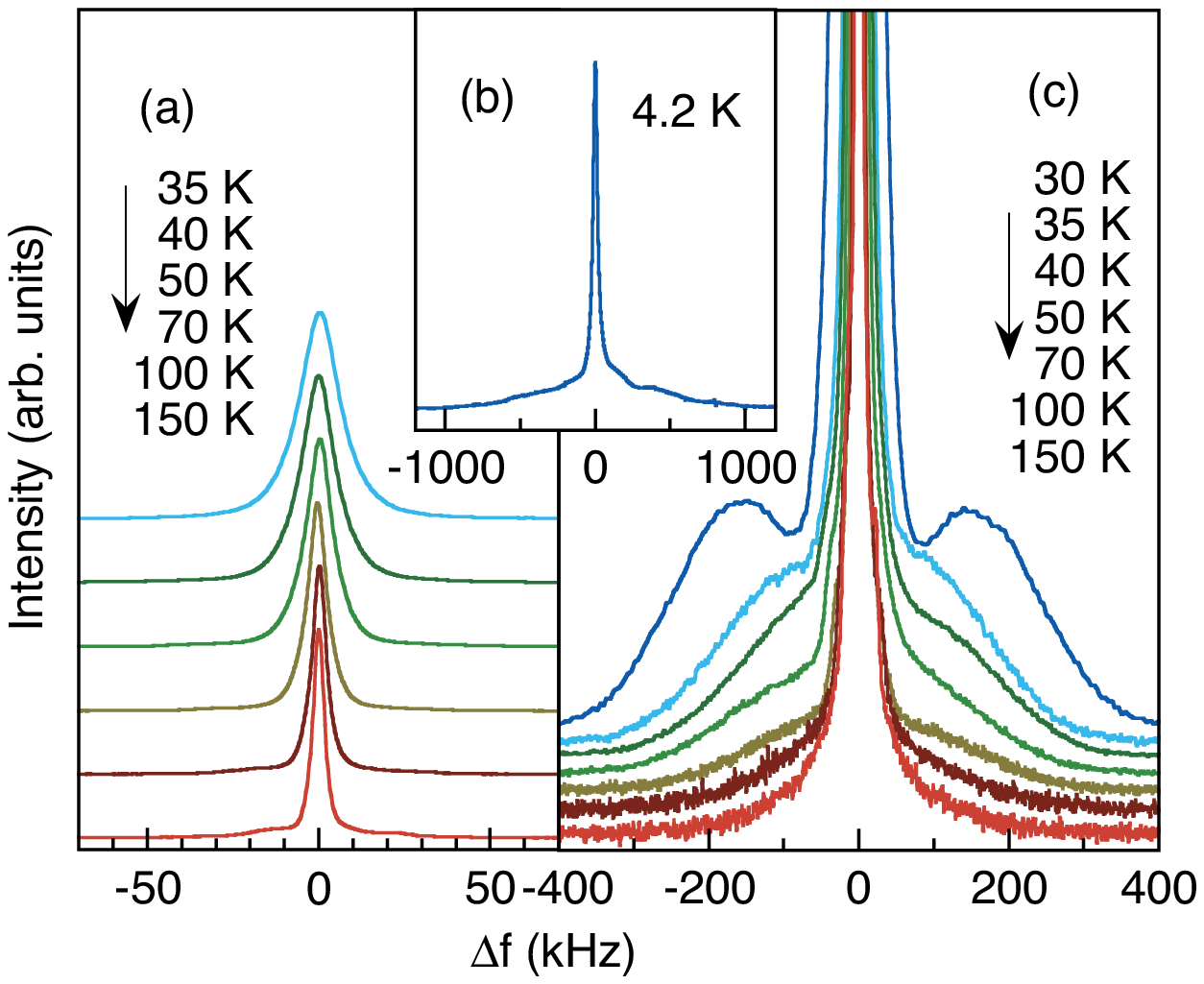}
 \end{center}
 \caption{\label{fig4} (color online).
$^{71}$Ga spectra measured above $T_\mathrm{S}$ (a) and below $T_\mathrm{S}$ (b), and magnified bottom structures above $T_\mathrm{S}$ (c). The data were collected in a field of 7.0071 T and plotted with respect to peak positions. The spectrum of (b) was obtained by summing Fourier transforms of the spin-echo signal accumulated at different frequencies.
}
 \end{figure}

Figure \ref{fig4}(a) shows $^{71}$Ga-NMR lines above $T_\mathrm{S}$.
A single symmetric line accords with the cubic symmetry of the Ga site ($4a$) above $T_\mathrm{S}$.
With decreasing $T$, the line  
is broadened.
We confirmed that the linewidth is magnetic in origin; the ratio of $^{69}$Ga and $^{71}$Ga linewidths is $T$-independent and close to $^{69}\gamma/^{71}\gamma = 0.7870$ ($\gamma$: gyromagnetic ratio) but different from $^{69}Q/^{71}Q = 1.60$ ($Q$: quadrupole moment). 
The $^{71}$Ga hyperfine shift, $^{71}K$, estimated from the peak position, is also plotted in Fig.\ \ref{fig3} \cite{note1}.
$^{71}K$ is in reasonable proportion to $\chi$ at least above $T_\mathrm{S}$ \cite{note3} in contrast to the case of $^{93}K$.
Below $T_\mathrm{S}$, $^{71}K$ drops suddenly, and approaches a constant value at low $T$, proving that the low-$T$ upturn in $\chi$ is extrinsic.
In addition, another component extended over 2 MHz appears on the foot of the center line as shown in Fig.\ \ref{fig4}(b).
This bottom component is most reasonably assigned to quadrupole satellites.
According to the lattice symmetry \cite{Jakob}, a single Ga site (nonaxial $2c$) exists only even below $T_\mathrm{S}$.
The unresolved satellites are reasonable if considerable asymmetry appears in $^{71}K$; note that the center line looks asymmetric.
Focusing on the bottom part of the spectrum, the $T$ variation is shown in Fig.\ \ref{fig4}(c).
Note that the quadrupole broadening already appears well above $T_\mathrm{S}$ in spite of the nominal cubic symmetry at the Ga site in the high $T$ state.
The result indicates that local environments around the Ga site already gets distorted well above $T_\mathrm{S}$.

\begin{figure}[tbh]
 \begin{center}
 \includegraphics[width=\linewidth]{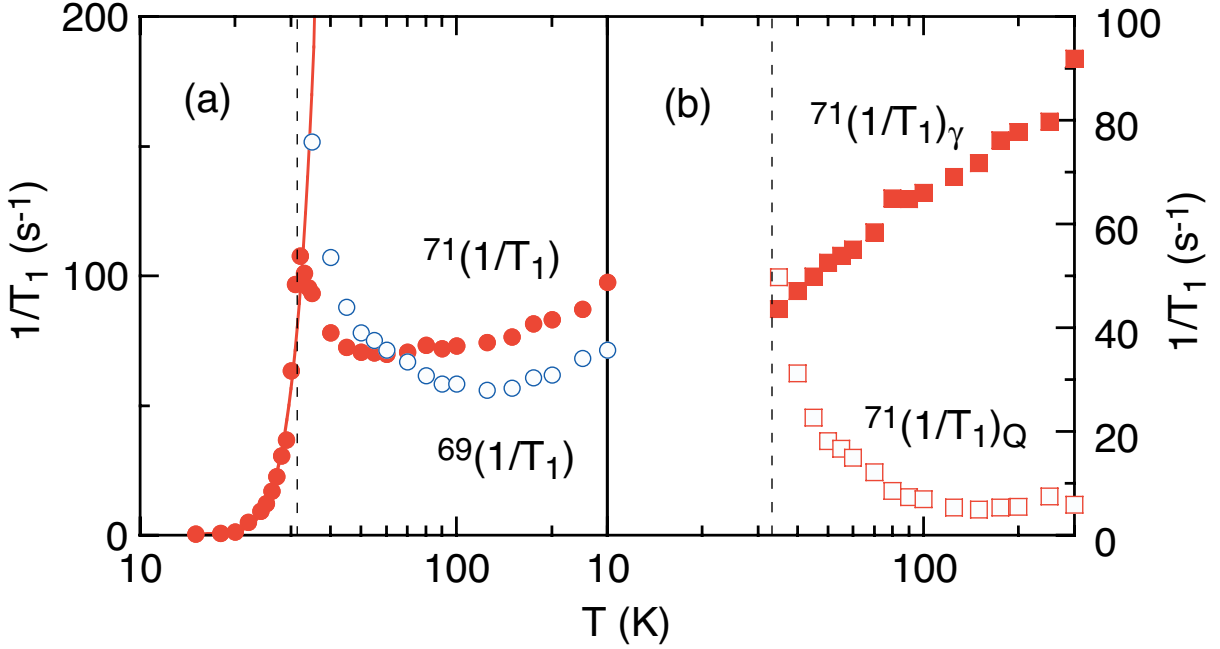}
 \end{center}
 \caption{\label{fig5} (color online).
(a) $T$ dependences of $1/T_1$ measured for $^{69}$Ga (open circles) and $^{71}$Ga (closed circles). Reliable $^{69}(1/T_1)$ was not obtained below $T_\mathrm{S}$ due to overlap of $^{93}$Nb and $^{69}$Ga signals.
The solid curve represents $\exp(-E_\mathrm{g}/k_\mathrm{B} T)$ with $E_\mathrm{g}/k_\mathrm{B} \simeq 190$ K
(b) $T$ dependences of quadrupole (open squares) and magnetic (solid squares) components of $^{71}(1/T_1)$.
 }
 \end{figure}

 Figure \ref{fig5}(a) shows $T$ dependences of $1/T_1$ for $^{69,71}$Ga.
 The recovery of nuclear magnetization fits well with a single exponential function well above $T_\mathrm{S}$, 
with two exponential functions expected for the $I=3/2$ nucleus with quadrupole interaction just above $T_\mathrm{S}$,
and 
with neither the single exponential nor the $I=3/2$ function below $T_\mathrm{S}$.
These facts are, as a whole, consistent with the observation in the static spectrum.
The last fact indicates the distribution of $1/T_1$, which is attributable to the appearance of hyperfine anisotropy.
Below $T_\mathrm{S}$, $1/T_1$ was estimated by applying a stretched exponential function $\exp(-(t/T_1)^{\beta})$ (obtained $\beta$ is nearly $T$ independent of $\sim$0.6).
 With decreasing $T$, $^{71}(1/T_1)$ once decreases to reach a minimum at $\sim$50 K, increases again to take a sharp peak at $T_\mathrm{S}$ and drops abruptly below $T_\mathrm{S}$.

Measurements of $1/T_1$ for different isotopes $^{69,71}$Ga give valuable information on spin and lattice dynamics.
As seen in Fig.\ \ref{fig5}(a), magnitudes of $^{69}(1/T_1)$ and $^{71}(1/T_1)$ are reversed at $\sim$60 K.
Because of $^{69}\gamma/^{71}\gamma <1$ and $^{69}Q/^{71}Q > 1$, this implies that both magnetic and electric quadrupole fluctuations contribute to $1/T_1$, and furthermore depend differently on $T$.
$1/T_1$ for the nucleus with $I$ and $Q$ is expressed as the superposition of magnetic and quadrupole terms, $1/T_1= (1/T_1)_{\gamma}+(1/T_1)_Q$.
Figure \ref{fig5}(b) shows $T$ dependences of separated contributions, estimated by solving simultaneous equations for $^{69}(1/T_1)$ and $^{71}(1/T_1)$ with use of relations $(1/T_1)_{\gamma} \propto \gamma ^2$ and $(1/T_1)_Q \propto Q^2$.
The magnetic component decreases monotonically, indicating no enhancement towards $T_\mathrm{S}$.
For the system with dynamically fluctuating local moments, we expect $T$-independent $1/T_1$.
The gradual decrease of $1/T_1$ is possibly ascribed to the development of spin-singlet correlations.
On the other hand, interestingly, the quadrupole contribution shows a critical behavior on approaching $T_\mathrm{S}$.
This enhancement indicates that structural fluctuations develop from $\sim$100 K, of the order of $3T_\mathrm{S}$.

ZF-$\mu$SR experiments detected no additional static field below $T_\mathrm{S}$. 
NMR spectral analyses are consistent with this observation.
These results prove unambiguously that the ground state is not static magnetic order.
This is in good contrast to the AF state in GeV$_4$S$_8$ established due to intercluster interactions \cite{Muller}.
As seen in Fig.\ \ref{fig3}, $^{71}K$ drops rapidly below $T_\mathrm{S}$, which is well fit by $\exp(-E_\mathrm{g}/k_\mathrm{B} T)$ with $E_\mathrm{g}/k_\mathrm{B} \simeq 220$ K.
As in Fig.\ \ref{fig5}(a), $^{71}(1/T_1)$ also shows a similar $T$ dependence with $E_\mathrm{g}/k_\mathrm{B} \simeq 190$ K.
Nearly the same estimation of the gap $\sim$200 K from both $K$ and $1/T_1$ indicates that the gap presents in spin excitations.
Hence we conclude that the ground state of GaNb$_4$S$_8$ is a $S = 0$ spin-singlet state.
The $T$ and field dependences of $\mu$SR-$\lambda$ also support this conclusion.

Characteristic precursory phenomena of the transition are summarized as 
(i) at the Nb site, the isotropic term of the local susceptibility $\chi _{\rm loc}$ looks suppressed from $\sim$$3T_\mathrm{S}$, and instead, large anisotropy appears, and
(ii) around the Ga site, in contrast to moderate magnetic fluctuations, both static and dynamic structural fluctuations are enhanced from $\sim$$3T_\mathrm{S}$.
As shown in Fig.\ \ref{fig1}, spin density distributions of HOMO are expected to be highly symmetric and asymmetric in the high-$T$ and low-$T$ states, respectively (similar discussions have been made on a 3d analog GaV$_4$S$_8$ \cite{Pocha2, Nakamura}).
On this basis, the above results are interpreted as follows; intracluster reconfiguration of Nb-4d cluster orbitals (i.e., mixing of the low-$T$ electronic state) already starts from $\sim$$3T_\mathrm{S}$, enhances intercluster structural instability, and finally induces cooperative long-range symmetry lowering at $T_\mathrm{S}$, which is coupled with the spin degree of freedom.

As the mechanism of the spin-singlet formation, spin compensation via intercluster interactions should be considered, because each Nb$_4$ cluster unit has one unpaired electron with $S =1/2$. 
The crystallographically unique Nb$_4$ cluster even below $T_\mathrm{S}$ \cite{Jakob} rules out charge disproportionation among clusters. 
According to the low-$T$ lattice symmetry \cite{Jakob}, all Nb$_4$ tetrahedra are elongated along one of the threefold axes $\langle 111 \rangle$ of the cubic lattice.
As a result, two Nb$_4$ tetrahedra come closer slightly along the tetragonal [110] or $[1\bar{1}0]$ direction.
Hence, if electron-lattice coupling is appreciable, it is reasonable to expect binding of two Nb$_4$ tetramers to a Nb$_8$ octamer (Nb$_4$--Nb$_4$ bond), just like the dimerization of the 1D spin chain.
We speculate that the transition at $T_\mathrm{S}$ is related with the characteristic electronic state of this material as a cluster Mott insulator \cite{Pocha, Jakob, Sieberer},
where the electron transfer is suppressed due to electron correlation, but is easily recovered by perturbation.
Simultaneously, as mentioned above, we should note that this transition is driven by the local JT-type instability within the cluster.
In this sense, the transition may be classified as the orbitally driven Peierls state \cite{Khomskii}, 
although intercluster spin-singlet correlation may also be developed from high-$T$ of the order of spin gap.
As a whole, the spin singlet state of GaNb$_4$S$_8$ is classified as one of exotic states found in frustrated lattices near metal-insulator boundaries;
the crystal structure of GaNb$_4$S$_8$ is closely related to the highly frustrated pyrochlore lattice, and simultaneously Nb$_4$ units arrange as a frustrated fcc lattice.

In conclusion, $\mu$SR and NMR experiments revealed that the ground state of GaNb$_4$S$_8$ is a nonmagnetic spin singlet with an excitation gap of $\sim$200 K. Taking into account the lattice symmetry below $T_\mathrm{S}$, formation of a Nb$_8$ octamer from two Nb$_4$ tetramers is suggested for the first time.
The transition is driven by the intracluster JT-type instability developed from high $T$ of the order of $3T_\mathrm{S}$.

This study was supported by Grant-in-Aid for Scientific Research on Priority Areas ``Novel States of Matter Induced by Frustration", 
and Grant-in-Aid for Young Scientists (Start-up) 
from MEXT, Japan.



\begin{thebibliography}{50}
\bibitem{note2} The values estimated from peak positions correspond to the component perpendicular to the principal axis, $K_\perp$.
Therefore the isotropic shift $K_\mathrm{iso}$ is expected to deviate more strongly from the Curie-Weiss behavior.
\bibitem{note3} The high-$T$ linear relations between $K$ and $\chi$ give the spin part in hyperfine coupling $^{93}A = -18.4$ and $^{71}A = 0.089$ T/$\mu_\mathrm{B}$, where we assumed the moment at each Nb atom and at each Nb$_4$ cluster for the cases of $^{93}$Nb and $^{71}$Ga, respectively.
\bibitem{note1} Below $T_\mathrm{S}$, the Ga center line shows anisotropic broadening; the shift at the peak position corresponds to $K_\perp$.
Note that the $T$ dependence of $^{71}K_\perp$ below $T_\mathrm{S}$, shown in Fig.\ \ref{fig3}, is stronger than that expected as $K_\mathrm{iso}$.
\end{thebibliography}
\end{document}